\documentclass[preprint, pra, aps, superscriptaddress, showpacs, groupaddress]{revtex4}

\usepackage{slashed}
\usepackage{graphicx}
\usepackage{subfigure}
\usepackage[usenames, dvipsnames]{color}
\usepackage{graphics}
\usepackage{hyperref}
\usepackage{bm}
\usepackage{amsmath}
\usepackage{color}
\usepackage{amsfonts}
\hypersetup{backref,
colorlinks=true,
linkcolor=cyan,
linktoc=page,
citecolor=cyan, 
urlcolor=cyan}

\everymath{\displaystyle}

\usepackage{epsfig}

\begin{document}

\title{Bose-Einstein condensation of photons in a plasma}

\author{J. T. Mendon\c{c}a}
\email{titomend@tecnico.ulisboa.pt}
\affiliation{Instituto de Plasmas e Fus\~ao Nuclear, Lisbon, Portugal}
\affiliation{Instituto Superior T\'ecnico, Universidade de Lisboa, Lisbon, Portugal}
\affiliation{Instituto de F\'isica, Universidade de S\~ao Paulo, S\~ao Paulo SP, Brazil}

\author{H. Ter\c{c}as}
\email{hugo.tercas@tecnico.ulisboa.pt}
\affiliation{Instituto de Plasmas e Fus\~ao Nuclear, Lisbon, Portugal}
\affiliation{Instituto de Telecomunica\c{c}\~oes, Lisbon, Portugal}

\begin{abstract}

We study the Bose-Einstein condensation of photons in a plasma, where we include the cases of both transverse photons and plasmons. We consider four-wave mixing processes of photon and plasmon  modes in a relativistic isotropic plasma to determine the coupling constant to lowest order. We further show that photon condensation is possible in an unbounded plasma because, in contrast with other optical media, plasmas introduce an effective photon mass. This guarantees the existence of a finite chemical potential and a critical temperature, which is calculated for both transverse photons and plasmons. By considering four-wave mixing processes, we derive the interactions between the photons in the condensate. We also study the elementary excitations (or Bogoliubov modes) of the condensed photon and plasmon gases, and determine the respective dispersion relations. Finally, we discuss the kinetics of photon condensation via inverse Compton scattering between the photons and the electrons in the plasma. 
\end{abstract}


\maketitle

\section{Introduction}

Despite the fact of Bose-Einstein statistics being first derived for photons \cite{bose}, the idea  of photon condensation itself has received much criticism from the physics community and therefore remained elusive. Two main reasons are at the origin of such reservations: first, photons are massless particles, which makes a vacuum in the lowest energy state (i.e. with infinite wavelength) impossible; second, it is difficult to implement a physical system where the number of photons is kept constant, a feature that is necessary to ensure a second-order phase transition. Strictly speaking, the existence of a chemical potential is crucial for photons to undergo Bose-Einstein condensation. Such problems do not arise - or, at least, are not that critical - in other bosonic systems. Indeed, successful experiments of Bose-Einstein condensates (BECs) with atoms \cite{anderson_1995, davis_1995, bradley_1995, donley_2002, durr_2004}, exciton-polaritons \cite{kasprzak_2006, deng_2010, guillaume_book} and, more recently, magnons \cite{slavin_2006, flebus_2016} are routinely reproduced nowadays. A way to circumvent the photon mass problem was first advanced in Refs. \cite{ciancaleoni_1997, chiao_1999}, where the photon mass appears as a consequence of the quantization of the electromagnetic modes along the axis of the cavity (actually, this mechanism also gives mass to the photon component of exciton-polaritons in semi-conductor microcavities \cite{guillaume_book, carusotto}). However, no thermalization process is discussed in Ref. \cite{chiao_1999}, and thus one can hardly imagine how condensation can indeed take place. \par

The recent observation of photon condensation in a Fabry-Perot cavity \cite{weitz_2010_1}, where the number of photons is nearly constant, has renewed the interest in the issue. In that work, the photons acquire an effective mass inside the cavity and pile up at the lowest level (i. e. the cavity cut-off frequency) due to the thermal equilibration resulting from the balance between absorption and re-emission events with the dye molecules. Since some losses are present (e.g. imperfections in the mirrors), a quasiconstant number of particles is achieved by pumping the cavity. Although a particle equilibration based on a delicate pump-loss balance does not necessarily lead to thermal equilibrium, a subsequent work by the same authors has produced compelling arguments favoring the observation of a genuine, second-order phase transition characterizing Bose-Einstein condensation of photons \cite{weitz_2010_2}. Another recent and appealing theoretical proposal, based on the concept of slow light \cite{slyusarenko_2008, slyusarenko_2009}, has been put forward by Boichenko and Slyusarenko \cite{boichenko_2015}, who considered the coexistence of photonic and atomic BECs. Moreover, a comprehensive theory of photon condensation in optical cavities in the presence of a medium has been developed by Kruchkov \cite{kruchkov_2014}. 
 
Another interesting physical medium where one could imagine photons to undergo Bose-Einstein condensation is the plasma. As it is well known, photons have an effective mass and an effective charge in a plasma \cite{mend2000}. This effective mass could reveal the existence of a vacuum process providing mass to the elementary particles, as first suggested by Anderson \cite{anderson}. This is the so-called Higgs mechanism \cite{higgs, englert, guralnik}, where an appropriate scalar field in vacuum replaces the free electrons in plasma. On the other hand, the existence of an effective charge could explain the ponderomotive force of laser pulses in a plasma, as well as the observed photon acceleration processes \cite{dias, book1}. As such, plasmas naturally solve the aforementioned photon mass problem and provide the finite chemical potential for BEC to occur. Contrary to the case of the experiments in optical cavities of Refs. \cite{ weitz_2010_1, weitz_2010_2}, photon condensation in a plasma is a bulk phenomenon, arising in homogeneous and unbounded systems. This is particularly relevante in the astrophysical context, where external trapping potentials are absent. Indeed, the possibility of photon BEC in a plasma was first considered by Zel'dovich and Levich in 1968 \cite{zeldovich}, in relation to the the famous distortion of the cosmic microwave background radiation through inverse Compton scatterin, the so-called Sunyaev-Zel'dovich effect \cite{sunyaev, birkinshaw}), but the subject did not received much attention since. However, reference should be made to more recent literature \cite{kuzmin, levan}, where nonlinear aspects of plasma-radiation interaction are treated. A comprehensive review on classical wave turbulence - a situation where phenomena closely related to photon condensation may arise - is provided in Ref. \cite{picozzi}. In any case, because a connection with fundamental aspects of Bose-Einstein condensation is still lacking, it is not surprising that the plasma case is not mentioned in the recent reviews \cite{carusotto}. \par

In the present work, we perform a comprehensive study of the different aspects of photon condensation in a plasma.  We provide a systematic quantum formulation of photon BECs in a relativistic plasma and a kinetic description of the Compton scattering, a number-conserving mechanism that allows photons to thermalize with the plasma electrons. We extend the discussion to the important case of {\it plasmons} (i. e. longitudinal photons), which has not been considered in the aforementioned works. The photon mass, chemical potential and critical temperature  are explicitly calculated. By considering four wave-mixing processes in the plasma, we are also able to determine the interaction constant between the BEC photons at the leading order. The low-frequency elementary excitations are consequently analyzed and comparison with the usual Bogoliubov spectrum of atomic BECs is established. In this paper, photons and plasmons will be refereed as {\sl quasi-particles}, as they represent collective excitations of the electron (and eventually ion) population mediated by the photon.  

This manuscript is organized as follows. In Sec. \ref{plasma_waves}, we state the wave equations governing the dynamics of photons and plasmons in isotropic plasmas. For generality, the plasmas are considered to be relativistic. In Sec. \ref{nonlinear_coupling}, we describe the nonlinear mode coupling between quasi-particle modes, and after discussing the cases of photons and plasmons separately, we propose a global formulation of the nonlinear evolution equations. The field quantization of the two kinds of plasma quasi-particles is performed in Sec. \ref{quantization}. In Sec. \ref{critical_temperature}, we derive the expressions for the critical temperature, using the standard Bogoliubov procedure \cite{pitaevskii, peth, book2}, and in Sec. \ref{elementary_excitations} we discuss the elementary excitations of the condensed quasi-particle gas. A wave-kinetic equation describing the thermal equilibration of photons via Compton scattering is studied in Sec. \ref{comptonization}. Finally, in Sec. \ref{conclusion}, some conclusions are stated.

\section{Plasma wave equations}
\label{plasma_waves}
Transverse photons and plasmons in isotropic plasmas are described by the vector potential ${\bf A}$ and the scalar potential $V$, respectively. In the Coulomb gauge, these potentials satisfy the following wave equations \cite{vieira_2015}
\begin{equation}
\left( \nabla^2 - \frac{1}{c^2} \frac{\partial^2}{\partial t^2} \right) {\bf a} = \frac{k_p^2}{\gamma} \frac{n}{n_0} {\bf u}_\perp ,
\label{2.1} 
\end{equation}
and
\begin{equation}
\nabla^2 \phi = \frac{e^2}{\epsilon_0 m_e  c^2} (n - n_0) .
\label{2.1b} 
\end{equation}
Here, $n$ is the electron density and $n_0$ is the equilibrium (ion) density. We have introduced the reduced vector and scalar potentials, ${\bf a} = e {\bf A} / m_e c$ and $\phi = e V/ m_ec^2$, where $e$ and $m$ are the electron charge and mass. We define the electron plasma wave number as $k_p = \omega_p / c$, where $\omega_p = \sqrt{e^2 n_0 / \epsilon_0 m_e}$ is the electron plasma frequency. Also, we have introduced the covariant velocity ${\bf u}$ and the relativistic gamma factor defined as
\begin{equation}
{\bf u} = \gamma \frac{{\bf v}}{c} \; , \quad \gamma = \sqrt{1 + u^2}.
\label{2.1c} 
\end{equation}
The transverse velocity component ${\bf u}_\perp$ satisfies the condition $\bm\nabla \cdot {\bf u}_\perp = 0$. The electron density $n$, and the velocity ${\bf u}$ can be determined by the relativistic electron fluid equations 
\begin{equation}
\frac{\partial n}{\partial t} + c \bm\nabla \cdot \left( \frac{n {\bf u}}{\gamma} \right) = 0,
\label{2.2} 
\end{equation}
and
\begin{equation}
\frac{\partial {\bf u} }{\partial t} + \frac{c {\bf u}}{\gamma} \cdot {\bf u} = - \frac{e}{m_e c} \left( {\bf E} + \frac{c {\bf u}}{\gamma} \times {\bf B} \right) - \frac {\bm\nabla P}{n m_e c}.
\label{2.2b} 
\end{equation}
The electron pressure $P$ can be related with the plasma density by using an appropriate equation of state, an the fields ${\bf E}$ and ${\bf B}$ can be in terms of the normalized potentials as
\begin{equation}
{\bf E} = - \frac{m_e c}{e} \left( \frac{\partial {\bf a}}{\partial t} + c \bm\nabla \phi \right) \; , \quad {\bf B} = \frac{m_e c}{e} \left( \bm\nabla \times {\bf a} \right) .
\label{2.2c} 
\end{equation}
We can linearize the above equations with respect to the perturbations. For transverse photons, we linearize with respect to ${\bf a}$ and assume that the density perturbations are equal to zero, $\tilde n \equiv n - n_0 = 0$. In contrast, for plasmons, we linearize with respect to $\phi$ and $\tilde n$ and neglect the vector potential perturbations ${\bf a} = 0$. We can then derive two distinct wave equations which can be cast into a single equation of the form
\begin{equation}
\left( \nabla^2 - \frac{1}{c_\sigma^2} \frac{\partial^2}{\partial t^2} \right) a_\sigma = \frac{\omega_p^2}{c_\sigma^2} a_\sigma \, .
\label{2.3} 
\end{equation}
Here, we have considered the three independent polarization (spin) states, $\sigma = \{0, \pm \}$. The state $\sigma = 0$ corresponds to longitudinal electrostatic waves, or plasmons, and $\sigma = \pm $ represent the two possible states of the transverse photons. For plasmons, we can use the scalar potential $a_0 \equiv \phi$, and the characteristic velocity is determined by the electron thermal velocity $c_0 \equiv S_e$, where $S_e = \sqrt{3 T_e / m_e}$. On the other hand, for the transverse photons, the normalized vector potential is $a_{\pm 1} = {\bf a}$ and the characteristic velocity is the light speed $c_{\pm 1} \equiv c$. For waves evolving in space and time as $a_\sigma \propto \exp (i {\bf k} \cdot {\bf r} - i \omega t)$, Eq. (\ref{2.3}) leads to the dispersion relations
\begin{equation}
\omega \equiv \omega_{\sigma, k} = \sqrt{\omega_p^2 + k^2 c_\sigma^2} = \frac{1}{\hbar} \sqrt{m_\sigma^2 c^4 + p^2 c^4} \, ,
\label{2.4} 
\end{equation}
where $\omega_p = \sqrt{e^2 n_0 / \epsilon_0 m_e}$ is the plasma frequency and $p = \hbar k$ is the quasi-particle momentum. For modes propagating along the $\bf \hat z$-direction, for plasmons we have ${\bf k} = k{\bf e}_z$, while for photons ${\bf a} \cdot {\bf e}_z = 0$. The two circular polarizations are then characterized by the unit polarization vectors ${\bf e}_k = {\bf a}_k / a_k$, such that ${\bf e}_k = ({\bf e}_x \pm {\bf e}_y) / \sqrt{2}$.\par

Relativistic effects can be introduced by dividing $\omega_p^2$ and $S_e^2$ by the relativistic gamma factor associated with thermal corrections, $\gamma_{\rm th} = \sqrt{1+S_e^2/c^2}$. For transverse electromagnetic modes, thermal effects are usually negligible. We notice that the dispersion relation (\ref{2.4}) allows us to introduce an effective mass for both plasmons and photons, which is given by
\begin{equation}
m_\sigma = \frac{\hbar \omega_p}{c_\sigma^2} \, .
\label{2.4b} 
\end{equation}
The existence of a finite mass is crucial for the existence of a finite critical temperature, allowing for Bose-Einstein condensation in an unbound plasma. This will be discussed in Sec. \ref{critical_temperature}. On the other hand, these quasi-particles are not free, but rather weakly interact with each other. As we show next, such an interaction is provided by wave-mixing, which is supported by the nonlinear properties of the medium. Similarly to the case of BECs in alkali atoms, interactions are at the origin of important quantum effects.

\section{Nonlinear mode coupling}
\label{nonlinear_coupling}
We now study the nonlinear evolution of wave excitations in a plasma. Returning to the basic equations, and making a perturbative analysis where the different modes are assumed to satisfy the linear dispersion relations (\ref{2.4}), we arrive at the nonlinear equation governing the weak nonlinear interaction (weak turbulence \cite{picozzi}) between the plasma quasi-particles. To clarify the discussion, we consider the cases of plasmons and photons separately.

\subsection{Plasmon coupling}

\begin{figure}[t!]
\includegraphics[scale=0.95]{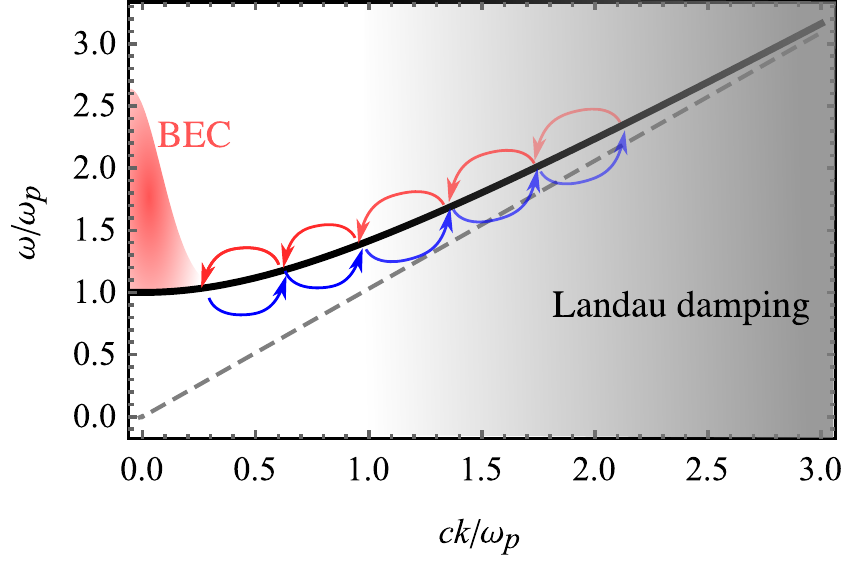}
\caption{(color online) Schematic representation of the Compton scattering leading to the thermalization of photons in a plasma. The photon dispersion relation in the plasma $\omega_k^2=\omega_p^2+c^2k^2$ (solid line) is compared to the photon dispersion in vacuum $\omega=ck$ (dashed line). The red downward arrows indicate the inverse Compton scattering for photons with initial temperature larger than the plasma temperature, leading to the piling of photons at the bottom of the dispersion (Bose-Einstein condensation). Compton processes with hot photons (blue upward arrows) leads to the spread of the photon distribution, thus preventing condensation. }
\label{fig_dispersion}
\end{figure}

For the electrostatic modes, the density perturbations $\tilde n = n -n_0$ are described by the following equation
\begin{equation}
\left( \frac{\partial^2}{\partial t^2} + \omega_p^2 - S_e^2 \nabla^2 \right) \tilde n = F_2 (\tilde n) + F_3 (\tilde n) \, ,
\label{3.1} 
\end{equation}
where the second and third order nonlinear terms are given by the auxiliary functions $F_2 (\tilde n)$ and $F_3 (\tilde n)$, respectively. We have
\begin{equation}
F_2 (\tilde n) = - \frac{e}{m_e} \bm \nabla \tilde n \cdot \bm \nabla V - \omega_p^2 \frac{\tilde n}{n_0} \tilde n \, ,
\label{3.1b} 
\end{equation}
and 
\begin{equation}
F_3 (\tilde n) = \bm \nabla \cdot \left[ {\bf v} \bm \nabla \cdot ( n  {\bf v}) + n {\bf v} \cdot \bm \nabla {\bf v}  \right] \, .
\label{3.1b} 
\end{equation}
We now introduce a Fourier expansion for the perturbed density $\tilde n \equiv \tilde n ({\bf r}, t)$, velocity ${\bf v} \equiv {\bf v} ({\bf r}, t)$ and electrostatic potential $V \equiv V ({\bf r}, t)$, of the form
\begin{equation}
\left( \tilde n, {\bf v}, V \right) = \int \left( \tilde n, {\bf v}, V \right)_k \exp( i {\bf k} \cdot {\bf r}) \frac{d {\bf r}}{(2 \pi)^3}.
\label{3.2}Ê
\end{equation}
As such, Eq. (\ref{3.1}) can be recast into the following wave-mixing equation
\begin{equation}
\left( \frac{\partial^2}{\partial t^2} + \omega_k^2 \right) \tilde n_k =  F_{2k} + F_{3 k} \, ,
\label{3.2b} 
\end{equation}
where the mode frequency is determined by $\omega_k^2 = \omega_p^2 + S_e^2 k^2$. The second and third order nonlinear terms are determined by
\begin{equation}
F_{2 k} = - \frac{\omega_p^2}{n_0} \sum_{k'} \left( 1 + \frac{{\bf k'} \cdot{\bf k''}}{k^{''2}} \right) \tilde n_{k'} \tilde n_{k''} \, ,
\label{3.2c} 
\end{equation}
where we have used ${\bf k''} = {\bf k} - {\bf k'}$, and 
\begin{equation}
F_{3 k} = - \frac{\omega_p^2}{n_0} \sum_{k' k''} \left\{ \left(\frac{{\bf k} \cdot {\bf k'}}{k^{'2}} \right) \left(\frac{{\bf k''} \cdot{\bf k'''}}{k^{'''2}} \right) + \left(\frac{{\bf k} \cdot{\bf k'^v}}{k'^{v2}} \right) \left(\frac{{\bf k'''} \cdot{\bf k'^v}}{k^{'''2}} \right) \right\}  \tilde n_{k'}  \tilde n_{k'''} \tilde n_{k'^v},
\label{3.3} 
\end{equation}
with ${\bf k'^v} = {\bf k} - {\bf k'} - {\bf k'''} = {\bf k''} - {\bf k'''}$. In order to derive Eq. (\ref{3.2c}) and (\ref{3.3}), we have followed the weak-turbulence approximation, which allows us to replace the quantities inside the nonlinear terms by their linear expressions, namely
\begin{equation}
{\bf v}_k = \frac{\omega_k}{k^2} \frac{\tilde n_k}{n_0} {\bf k} \, , \quad V_k = - \frac{e \tilde n_k}{\epsilon_0 k^2} \, .
\label{3.3b}
\end{equation}
In the above expressions, the summations are to be replace by integrals over the spectrum
\begin{equation}
\sum_k \rightarrow \int \frac{d {\bf k}}{(2 \pi)^3} \, , \quad \sum_{k k'} \rightarrow \int \frac{d {\bf k}}{(2 \pi)^3} \int \frac{d {\bf k'}}{(2 \pi)^3} \, .
\label{3.3c} 
\end{equation}
We notice that the nonlinear terms $F_{2k}$ and $F_{3k}$ represent a resonant coupling between three- and four-wave modes inside the plasmon spectrum. The first ones were established assuming that $\omega_k - \omega_{k'} \pm \omega_{k''} = 0$, and the second ones are satisfied for $\omega_k - \omega_{k'} =\omega_{k'''} - \omega_{k'^v} $. It can be readily seen that three-wave mixing is suppressed inside the plasmon spectrum, if the three modes involved are close to the plasma frequency $\omega_p$ where condensation is expected to arise (see Fig. \ref{fig_dispersion} for a schematic illustration). In fact, three-wave mixing would involve modes in the linear part of the spectrum, $\omega\sim ck$, which are strongly damped via photon Landau damping \cite{bingham_1997, mendonca_2006}. In contrast, four-wave mixing is possible in the same region of the spectrum. In what follows, we will therefore neglected the second-order term $F_{2 k}$. Thus, we can now rewrite Eq. (\ref{3.2b}) in terms of the (normalized) potential perturbations $\phi_k$ as
\begin{equation}
\left( \frac{\partial^2}{\partial t^2} + \omega_k^2 \right) \phi_k =  - \sum_{k' k'''} C_0 ({\bf k}, {\bf k'},{\bf k'''}) \phi_{k'} \phi_{k'''} \phi_{k'^v} \, ,
\label{3.4} 
\end{equation}
with the plasmon coupling coefficients
\begin{equation}
C _0({\bf k}, {\bf k'},{\bf k'''}) =  \frac{m_e c^2}{e \mu_0} \left[ \left(\frac{{\bf k} \cdot {\bf k'}}{k^{'2}} \right) \left(\frac{{\bf k''} \cdot{\bf k'''}}{k^{'''2}} \right) + \left(\frac{{\bf k} \cdot{\bf k'^v}}{k'^{v2}} \right) \left(\frac{{\bf k'''} \cdot{\bf k'^v}}{k^{'''2}} \right) \right] \left( \frac{k' k''' k'^v}{k} \right)^2 \, ,
\label{3.4b} 
\end{equation}
where $\mu_0 = 1 / c^2 \epsilon_0$. These equations will allow us to study the main properties of plasmon condensation.

\subsection{Photon coupling}

We now turn to the case of photons in a plasma, by going back to Eq. (\ref{2.1}) and performing the same perturbative analysis. We notice that, in the case of photons, the linear plasma response implies that $\tilde n = 0$ and ${\bf u}_\perp = {\bf a}$. This means that the main source of nonlinearity is the relativistic gamma factor $\gamma$. We can then rewrite Eq. (\ref{2.1}) into a more amenable form
\begin{equation}
\left( \nabla^2 - \frac{1}{c^2} \frac{\partial^2}{\partial t^2} \right) {\bf a} = \frac{k_p^2}{\gamma} \, {\bf a} \, ,
\label{3.5} 
\end{equation}
where the relativistic factor, neglecting thermal corrections, can be written as $\gamma = \sqrt{1 + a^2}$. We now proceed to a Fourier decomposition of the photon field, ${\bf a} \equiv {\bf a} ({\bf r}, t) = \sum_k {\bf a}_k \exp (i {\bf k} \cdot {\bf r})$, which, assuming circular polarization, yields
\begin{equation}
\gamma = \sqrt{\gamma_a^2 + S} \, , \quad \gamma_a = \sqrt{1 + \sum_k \left| a_k \right|^2} \, 
\label{3.6} 
\end{equation}
with
\begin{equation}
S = \sum_{k \neq k'} \left({\bf a}_k \cdot {\bf a}_{k'} \right) \exp \left[ i ({\bf k} - {\bf k'}) \cdot {\bf r} \right].
\label{3.6b} 
\end{equation}
In our description, the quantity $\gamma_a$ is not time-dependent. As a matter of fact, if linear polarization was considered instead, $\gamma_a$ would strongly evolve in time and would lead to the generation of a large spectrum of harmonics \cite{book1}. By expanding the factor $1/ \gamma$ on the r.h.s. of Eq. (\ref{3.5}), we can obtain \cite{vieira_2015}
\begin{equation}
\left( \nabla^2 - \frac{1}{c^2} \frac{\partial^2}{\partial t^2} - \frac{k_p^2}{\gamma_a}  \right) {\bf a} = \frac{k_p^2}{\gamma_a} {\bf a} \sum_\ell C_\ell \left(\frac{S}{\gamma_a^2} \right)^\ell\,  .
\label{3.7} 
\end{equation}
This expansion is valid for arbitrary intensities, because the inequality $ (S / \gamma_a^2) \leq 1$ is always verified. We should also notice that the first coefficients in this expansion are $C_1 = - 1/2$, and $C_2 =(2/3)^3 < C_1$. Due to the increasing powers of $\gamma_a^{-2 \ell}$, the higher order terms can be neglected, and only the first nonlinear term, corresponding to $\ell= 1$, will be retained. Using ${\bf a}_k = a_k {\bf e}_k$, where the unit polarization vectors satisfy $({\bf e}_k^* \cdot {\bf e}_k) = 1$ and the described circularly polarized states, we can recast Eq. (\ref{3.7}) into
\begin{equation}
\left( \frac{\partial^2}{\partial t^2} + \omega_k^2 \right) a_k =  - \sum_{k' k''} C_\pm ({\bf k}, {\bf k'}, {\bf k''}) a_{k'} a_{k''} a_{k'''},
\label{3.8} 
\end{equation}
where we have used the dispersion relation $\omega_k^2 = (\omega_p^2 / \gamma_a) + k^2 c^2$, and defined ${\bf k'''} = {\bf k} + {\bf k''} - {\bf k'}$. The nonlinear coupling coefficients are determined by 
\begin{equation}
C _\pm({\bf k}, {\bf k'}, {\bf k''}) = \frac{k_p^2}{2 \gamma_a^3} ({\bf e}_{k'} \cdot {\bf e}_{k''}) ({\bf e}_k^* \cdot {\bf e}_{k'''}).
\label{3.8b} 
\end{equation}

\section{Field quantization}
\label{quantization}
We are now ready to establish the quantum description of the plasma modes, which are classically governed by Eqs. \eqref{3.4} and \eqref{3.8b} . This will then be used to determined the basic properties of a plasma condensate. For that purpose, we start by introducing the total energy associated with the two quasi-particle populations. We use
\begin{equation}
W_\sigma = \delta_\sigma \sum_k W_{\sigma, k} \, , 
\label{4.1} 
\end{equation}
where $\delta_\sigma$ accounts for both plasmon and photon polarizations ($\delta_0 = 1$ and $\delta_\pm = 2$). The energy density per mode is determined by
\begin{equation}
W_{\sigma, k} = \frac{\epsilon_0}{2} \left| E_k \right|^2 \left( \frac{\partial \epsilon_\sigma (\omega)}{\partial \omega} \right) \, .
\label{4.1b} 
\end{equation}
For plasmons, the energy is equally divided between electrostatic energy and kinetic energy of the electrons. In contrast, for photons, the energy is equally divided between electric and magnetic energy. Here, we have defined the dielectric functions as
\begin{equation}
\epsilon_0 (\omega) = 1 -  \frac{\omega_p^2}{\omega^2} - \frac{k^2 S_e^2}{\omega^2}, \quad \epsilon_\pm  = 1- \frac{\omega_p^2}{\gamma_a^2 \omega^2}- \frac{k^2 c^2}{\omega^2} \, .
\label{4.2} 
\end{equation}
Notice that, in isotropic plasmas, the two polarizations states $\sigma = \pm 1$ are degenerate. The same would not true for a magnetized plasma. Using these expressions in Eq. (\ref{4.1b}), we get
\begin{equation}
W_{\sigma, k} = \frac{\epsilon_0}{2}  f_\sigma (\omega) \left( \frac{m_e c}{e} \right)^2  \left| a_{\sigma, k} \right|^2 ,
\label{4.2b} 
\end{equation}
with the auxiliary functions defined as
\begin{equation}
f_{0} (\omega) = 2 k^2 c^2 \, , \quad 
f_{\pm } (\omega) =  \omega^2 \left(1 + \frac{\omega_p^2}{\gamma_a^2 \omega^2} \right) \, .
\label{4.2c} 
\end{equation}
In a first step, we neglect the interaction energy due to the nonlinear mode coupling. This will be considered later. We now introduce the field operators
 \begin{equation}
 \hat a_{\sigma, k} ({\bf r}, t) = \beta_{\sigma, k} \left[ \hat b_k ( t) e^{i {\bf k} \cdot {\bf r}} + \hat b_k^\dag ( t ) e^{- i {\bf k} \cdot {\bf r}} \right]_\sigma \, ,
\label{4.3} 
\end{equation}
where $\beta_{\sigma, k}$ are normalization constants, to be defined below, and the operators $(\hat b_k, \hat b_k^\dag)$ satisfy the usual boson commutation relations
\begin{equation}
\left[ \hat b_k \, , \hat b_{k'}^\dag \right] = \delta ({\bf k} - {\bf k'}) \, , \quad 
\left[ \hat b_k \, , \hat b_{k'} \right] = \left[ \hat b_k^\dag \, , \hat b_{k'}^\dag \right] = 0 \, . 
\label{4.3b} 
\end{equation}
It can easily be seen that 
\begin{equation}
\left| \hat a_{\sigma, k} \right|^2 = \left| \beta_{\sigma, k} \right|^2 \left(\hat b_k^\dag \hat b_k + \hat b_k \hat b_k^\dag \right)_\sigma \,.
 \label{4.4} 
 \end{equation}
Comparing this expression with that of the classical energy density in Eq. (\ref{4.2b}), we can define the Hamiltonian $\hat H_{\sigma, k}$ for which the classical energy would be its expectation value, $W_{\sigma, k} = \langle \hat H_{\sigma, k} \rangle$, yielding
\begin{equation}
\hat H_{\sigma, k} = \frac{ \epsilon_0}{2} f_\sigma (\omega) \left| \beta_{\sigma, k} \right|^2 \left(\hat b_k^\dag \hat b_k + \hat b_k \hat b_k^\dag \right)_\sigma.
\label{4.4b} 
\end{equation}
 Using the normalization factor
 \begin{equation}
\beta_{\sigma, k} = \sqrt{\frac{\hbar \omega_{\sigma, k}}{\epsilon_0 f_\sigma (\omega)}} \, ,
\label{4.4c} 
\end{equation}
we can rewrite the free Hamiltonian in Eq. \eqref{4.4b} as
\begin{equation}
\hat H_{\sigma, k} =  \hbar \omega_{\sigma, k} \left(\hat b_k^\dag \hat b_k + \frac{1}{2} \right)_\sigma \, .
\label{4.5} 
\end{equation}
From here, we can establish the Heisenberg equations for the temporal evolution of the mode operators, using the general expression
\begin{equation}
\frac{d \hat b_k}{d t} = \frac{1}{i \hbar} \left[ \hat b_k , \hat H_{\sigma, k} \right] \, , \quad
\frac{d \hat b_k^\dag}{d t} = - \frac{1}{i \hbar} \left[ \hat b_k^\dag , \hat H_{\sigma, k} \right] 
\, .
\label{4.5b} \end{equation}
Using the above commutation relations, we can easily get
\begin{equation}
\frac{d \hat b_k}{d t} = - i \omega_k \hat b_k \, , \quad
\frac{d \hat b_k^\dag}{d t} = i \omega_k \hat b_k^\dag \, ,
\label{4.6} \end{equation}
with the obvious solutions
\begin{equation} 
\hat b_k ( t ) = \hat b_k ( 0 ) \exp (- i \omega_k t) \, , \quad
\hat b_k^\dag ( t ) = \hat b_k^\dag ( 0 ) \exp ( i \omega_k t) \, .
\label{4.6b} \end{equation}
These solutions are only valid in the linear regime, when mode-mode coupling is disregarded. Generalization to the nonlinear regime implies a similar calculation for an interaction Hamiltonian $\hat H_{\sigma, \rm int}$, such that the total Hamiltonian becomes
\begin{equation}
\hat H = \sum_{\sigma, k} \hat  H_{\sigma, k} + \hat H_{\sigma, {\rm int}}
\label{4.6c} 
\end{equation}
We proceed below to such a calculation. Dropping the polarization state $\sigma$ for simplicity, we can then refine eqs. (\ref{4.6}) replacing them by
\begin{equation}
\frac{d \hat b_k}{d t} = - i \omega_k \hat b_k + \frac{1}{i \hbar} \left[ \hat b_k , \hat H_{\rm int} \right] \, , \quad
\frac{d \hat b_k^\dag}{d t} = i \omega_k \hat b_k^\dag - \frac{1}{i \hbar} \left[ \hat b_k^\dag , \hat H_{\rm int} \right] \, ,
\label{4.7} \end{equation}
By iterating the latter, we can obtain
\begin{equation}
\frac{d^2 \hat b_k}{d t^2} + \omega_k^2 \hat b_k = - \frac{\omega_k}{\hbar} \left[ \hat b_k , \hat H_{\rm int} \right] \, , 
\label{4.7b} 
\end{equation}
with a similar equation for $\hat b_k^\dag$. Comparing this with the classical mode equations above, we can then construct the appropriate interaction Hamiltonian for each polarization mode as
\begin{equation}
H_{\sigma, {\rm int}} = \frac{1}{2} \sum_{{\bf k}, {\bf p}, {\bf q}} \mathcal{C}_{\sigma} ({\bf k}, {\bf p}, {\bf q}) \hat b_{\sigma, {\bf k} + {\bf p}}^\dag \hat b_{\sigma, {\bf q} - {\bf p}}^\dag \hat b_{\sigma, \bf k} \hat b_{\sigma, \bf q}  \, .
\label{4.8} 
\end{equation}
Here, the summation indices were rearranged for future convenience. The corresponding mode coupling coefficients read
\begin{equation}
\mathcal{C}_{\sigma} ({\bf k}, {\bf p}, {\bf q}) = \alpha_\sigma C_\sigma ({\bf k} + {\bf p}, {\bf q} - {\bf p}, {\bf q}),
\label{4.8b} 
\end{equation}
where we have defined
\begin{equation}
\alpha_{0} = \frac{\omega_k}{\hbar} \, , \quad
\alpha_{ \pm } = \frac{1}{\epsilon_0}  \left( \frac{m_e c}{e} \right)^2 \left(1 + \frac{\omega_p^2}{\gamma_a^2 \omega_k^2} \right)^{-1}.
\label{4.8c} 
\end{equation}
In the case of photons, we explicitly obtain
\begin{equation}
\mathcal{C}_{\pm } ({\bf k}, {\bf p}, {\bf q}) = \frac{m_e^2}{e \epsilon_0 \gamma_a} 
\frac{({\bf e}_{{\bf k} + {\bf p}} \cdot {\bf e}_{{\bf q} -{\bf p}})({\bf e}_{\bf k}^* \cdot {\bf e}_{\bf q})}{1 + \gamma_a^2 \omega_k^2 / \omega_p^2}.
\label{4.9} 
\end{equation}
The Hamiltonian operator in Eq. (\ref{4.8}) describes a quantum gas of plasmons or photons, where these quasi-particles weakly interact through four-wave mixing processes. 

\section{Critical temperature}
\label{critical_temperature}

We can now consider such a bosonic gas of quasi-particles, in thermal equilibrium, and establish the main properties of Bose-Einstein condensation. We assume that the number of quasi-particles is conserved, which agrees with most experimental conditions for both photons and plasmons. However, in the general case, the plasma should be considered as an open system, where bosons are injected from an external source in order to compensate the small but unavoidable losses. For plasmons, losses are mainly due to electron Landau damping, and for photons due to inverse bremsstrahlung. In any case, for a condensed gas with $\omega\simeq \omega_p$, these loss mechanisms are strongly suppressed. In order to determine the critical temperature, below which condensation is achieved, we assume that the expectation value for the mode number operators satisfies the Bose-Einstein distribution
\begin{equation}
N_\sigma ({\bf k}) \equiv \langle \hat b_k^\dag \hat b_k \rangle_\sigma  = \frac{1}{\exp (\hbar \omega_{\sigma, k} - \mu_\sigma) / T_\sigma - 1} \, ,
\label{5.1} 
\end{equation}
where the temperature of the photon gas $T_\sigma$ can be different from the plasma temperature, as shown below. Here, the mode frequency $\omega_{\sigma, k}$ still satisfies the dispersion relation (\ref{2.4}), and $\mu_\sigma$ is the chemical potential to be determined below. Using the quasi-particle momentum ${\bf p} = \hbar {\bf k}$, we can also write
\begin{equation}
N_\sigma ({\bf p})  = \frac{1}{\exp ( \epsilon_{\sigma, p} - \mu_\sigma) / T_\sigma - 1} \, ,
\label{5.1b} 
\end{equation}
where $\epsilon_{\sigma,p} = \hbar \omega_{\sigma, k}$ is the quasi-particle energy, as defined by
\begin{equation}
\epsilon_{\sigma,p} = \sqrt{m_\sigma^2 c^4 + p^2 c_\sigma^4} \, , \quad m_\sigma = \frac{\hbar \omega_p}{c_\sigma^2} \, .
\label{5.1c} 
\end{equation}
Notice that the equivalent masses for plasmons and photons are quite different, 
\begin{equation}
m_{0} = \frac{\hbar \omega_p}{S_e^2} = \frac{\hbar}{\lambda_{De} S_e} \, , \quad m_{\pm } = \frac{\hbar \omega}{c^2 \sqrt{\gamma_a}},
\label{5.2} 
\end{equation}
with $\lambda_{De} = S_e / \omega_p$ denoting the electron Debye length. The photon effective mass $m_{\pm }$ is much smaller than the plasmon effective mass $m_0$ (actually, $m_\pm/m_0=S_e^2 / (c^2 \sqrt{\gamma_a}) \ll 1$). The two polarization states usually coexist in equal parts in a plasma, so we should use, for the total number of quasi-particles
\begin{equation}
n_\sigma = 2^\delta \int N_\sigma ({\bf k}) \frac{d {\bf k}}{(2 \pi)^3} \rightarrow 2^\delta \sum_k N_\sigma ({\bf k}) \, .
\label{5.2b} 
\end{equation}
We can also define the {\it quasi-particle fugacity}
\begin{equation}
z_\sigma = \exp \left[ \left( \mu_\sigma - m_\sigma c_\sigma^2 \right) / T_\sigma \right] \, .
\label{5.3} 
\end{equation}
Replacing this in Eqs. (\ref{5.1b})-(\ref{5.2b}), we obtain
\begin{equation}
n_\sigma = 2^\delta \int  \frac{1}{z_\sigma^{-1} \exp ( w_\sigma / T_\sigma) - 1} \frac{d {\bf k}}{(2 \pi)^3} \, ,
\label{5.3b} 
\end{equation}
with the quasi-particle kinetic energy defined by $w_\sigma = \epsilon_\sigma - m_\sigma c_\sigma^2$. The critical temperature $T_{c, \sigma}$ for quasi-particle condensation is determined by the condition $n_\sigma = 0$, which implies that $w_\sigma=\mu_\sigma - m_\sigma c_\sigma^2=0$. Condensation is therefore defined at the bottom of the dispersion $p = 0$, which corresponds to the following value for the chemical potential at critical temperature
\begin{equation}
\mu_\sigma (T_{c, \sigma}) = m_\sigma c_\sigma^2 = \frac{\hbar \omega_p}{\sqrt{\gamma_\sigma}} \, ,
\label{5.4} 
\end{equation}
and the corresponding critical fugacity $z_\sigma (T_{c, \sigma}) = 1$. Assuming isotropic quasi-particle spectrum inside the plasma, we can make the replacement $d {\bf p} \rightarrow 4 \pi p^2 d p$ in the integral and obtain
\begin{equation}
n_\sigma = \frac{2^\delta}{4 \pi^2 \lambda_\sigma^3} \int_0^\infty  \frac{\sqrt{x} \, d x}{\exp (a_\sigma \sqrt{1 + x} - a_\sigma) - 1} \, ,
\label{5.4b} 
\end{equation}
where we have introduced the following dimensionless quantities
\begin{equation}
x = \frac{p^2}{m_\sigma^2 c_\sigma^2} \, , \quad a_\sigma = \frac{m_\sigma c_\sigma^2}{T_{c, \sigma}} \, , \quad \lambda_\sigma = \frac{\hbar}{m_\sigma c_\sigma} \, .
\label{5.5} 
\end{equation}
We should also note that
\begin{equation}
a_\sigma = \frac{\hbar \omega_p}{T_{c, \sigma}} \sqrt{\gamma_\sigma} \, , \quad \lambda_\sigma = \lambda_C \frac{m}{m_\sigma} \frac{c}{c_\sigma} \, ,
\label{5.5b} 
\end{equation}
where $\lambda_C = \hbar / m c$ is the Compton wavelength. After some algebraic manipulation, Eq. (\ref{5.4b}) can be transformed into
an infinite series, of the form
\begin{equation}
n_\sigma = \frac{2^\delta}{2 \pi^2 \lambda_\sigma^3} \sum_{\nu = 1}^\infty  \frac{\exp (\nu a_\sigma)}{\nu a_\sigma} K_2 (\nu  a_\sigma) \, ,
\label{5.6} 
\end{equation}
where $K_2 (x)$ is the modified Bessel function of the second kind, also known as Macdonald function. For photons, this coincides with Eq. (16) of Ref. \cite{levan}. In the low density (or strongly relativistic) limit, such that $a_\sigma \gg 1$, or $\gamma_\sigma \gg (\hbar \omega_p / T_{c, \sigma})^2$, we can use the approximation of $K_2 ( z )$ for $z$ large (see Ref. \cite{abramovich}, Eq. 9.7.2), which yields
\begin{equation}
n_\sigma \simeq \frac{2^\delta}{(2 \pi)^{3/2} \lambda_\sigma^3} \sum_{\nu = 1}^\infty  \frac{1}{(\nu a_\sigma)^{3/2}} \left[1 + \frac{15}{8}\frac{1}{\nu a_\sigma} + \cdots \right] \, .
\label{5.6b} 
\end{equation}
To the leading order, we obtain the critical temperature as
\begin{equation}
T_{c, \sigma} \simeq \pi  \frac{\hbar^2}{m_\sigma} n_\sigma^{2/3} \, .
\label{5.7} 
\end{equation}
Apart from a meaningless factor of order one, this is the same result for an ideal Bose gas with mass $m$ and density $n$ \cite{pitaevskii, peth}. The only difference here is the inclusion of the effective mass $m_\sigma$ and the quasi-particle density, which is typically the total electromagnetic energy density divided by the plasmon energy, $n_\sigma \sim \sum_k W_{{k}, \sigma} / \hbar \omega_p$. On the other hand, when we compare photons and plasmons, we notice that the photon effective mass is smaller, by a factor $(S_e^2 / c^2)$, than the mass of plasmons. This means that the critical temperature is much higher, by the same factor, and for the same total electromagnetic energy. We conclude that it is much easier, at least in principle, to condensate photons than plasmons in a plasma.

\section{Elementary excitations}
\label{elementary_excitations}

In order to study the elementary excitations on top the plasmon/photon condensate, we consider the total Hamiltonian in Eq \eqref{4.6c}, introduce the many-body Hamiltonian
\begin{equation}
\mathcal{\hat H}_\sigma = \hat H_\sigma-\mu_\sigma \hat N
\end{equation}
and assume the photon and plasmon gas independently. Following the Bogoliubov prescription, we write the quasi-particle number operator as
\begin{equation}
\hat N \equiv \sum_k \hat b_k^\dag \hat b_k \simeq N_{0, \sigma} + \sum_k' \hat b_k^\dag \hat b_k,
\label{6.1} 
\end{equation}
where $N_{0,\sigma}$ is the number os condensed quasi-particles, and the sum $\sum'$ excludes the condensed state ${\bf k} = 0$. Replacing $\hat b_0^\dag$ and  $\hat b_0$ by $\sqrt{N_{0, \sigma}}$ in Eq. \eqref{4.6c}, we can divide the Hamiltonian as
\begin{equation}
\mathcal{\hat H}_\sigma = \mathcal{\hat H}_{0, \sigma} + \mathcal{\hat H}_{1, \sigma} + \mathcal{\hat H}_{2, \sigma} \, ,
\label{6.1b} 
\end{equation}
with the lowest order term given by
\begin{equation}
\mathcal{\hat H}_{0, \sigma} = \left(\hbar \omega_p-\mu_\sigma\right) N_{0, \sigma} + \frac{1}{2} \mathcal{C}_\sigma (0) N_{0, \sigma}^2 \, ,
\label{6.2} 
\end{equation}
the first order term is
\begin{equation}
\mathcal{\hat H}_{1, \sigma} = \frac{1}{2} N_{0, \sigma} \sum_k' \mathcal{C}_\sigma ({\bf k}) \left\{ \left( \hat b_k^\dag \hat b_{- k}^\dag + \hat b_{- k} \hat b_k \right) + \left( \hat b_k^\dag \hat b_k + \hat b_{- k}^\dag \hat b_k \right) \right\}_\sigma 
+\sum_k' \epsilon_{k,\sigma} \hat b_k^\dag \hat b_k  \, ,
\label{6.2b} 
\end{equation}
with $\epsilon_{k,\sigma}=\hbar\omega_{k,\sigma}-\mu_\sigma$ and, finally, the second-order term
\begin{equation}
\mathcal{\hat H}_{2, \sigma} = \frac{1}{2} \sqrt{N_{0, \sigma}} \sum_{k,p}' \mathcal{C}_\sigma ({\bf k}, {\bf p}) \left\{ \left( \hat b_{k+ p}^\dag \hat b_k \hat b_p + \hat b_{k+p}^\dag \hat b_{-k}^\dag \hat b_p \right)_\sigma +  \frac{1}{2} \sum_{k,p,q}' \mathcal{C}_\sigma ({\bf k}, {\bf p}, {\bf q}) \left( \hat b_{k+p}^\dag \hat b_{q-p}^\dag \hat b_k \hat b_q \right)_\sigma\right\} \, .
\label{6.2c} 
\end{equation}
For a large number of condensed quasi-particles, $N_{0, \sigma} \gg 1$, we have $\mathcal{\hat H}_{0,\sigma} \gg \mathcal{\hat H}_{1,\sigma} \gg \mathcal{\hat H}_{2,\sigma}$, where the zero order Hamiltonian describes the ground state, and the first order term contains the main contributions from both quantum and thermal fluctuations. In order to appropriately describe such fluctuations, we should retain the expansion up to $\mathcal{\hat H}_{1, \sigma}$. This term can be more appropriately written as
\begin{equation}
\mathcal{\hat H}_{1,\sigma} = \sum_k' \left\{ \left(\epsilon_k + g_{\sigma, k} N_0 \right) \hat b_k^\dag \hat b_k + \frac{1}{2} g_{\sigma, k}N_0 \left( \hat b_k^\dag \hat b_{-k}^\dag \hat b_{-k} \hat b_k \right) \right\}_\sigma \, ,
\label{6.3} 
\end{equation}
which represents the free quasi-particle energy, and defined the interaction parameter $g_{\sigma, k} = \mathcal{C}_\sigma ({\bf k}) / N_{0, \sigma}$. We notice that, for both cases of plasmons and photons this quantity is positive, $g_{\sigma, k} > 0$, which means that the nonlinear interactions inside the plasmon or the photon gas are always repulsive. We therefore concede that there is no plasmon/photon collapse after condensation. \par

At this point, it should be noticed that the term {\sl plasmon collapse} has occasionally been used in the literature \cite{collapse}, but this is due to an attractive plasmon interaction mediated by phonons. For plasmons, such attractive interactions would appear as the three-wave mixing processes described above, if one of the mixing waves is an ion acoustic wave.
This has been ignored here, because we have focused on the Bose-Einstein condensation of isolated plasmon (or photon) gas, disregarding the lower frequency oscillations associated with the ion motion. As discussed in Sec. II, only the high frequency oscillations associated with the electron plasma response were retained, and the ions were assumed as an immobile background. This means that the concept of plasmon collapse does not apply to Bose-Einstein condensation of a plasmon gas. However, the ion motion will eventually play a role in the plasmon thermal fluctuations, and its influence of the dynamics of the BEC should be addressed in a separate publication.\par

Going back to the Hamiltonian in Eq. (\ref{6.3}), we can simplify it by evoking a Bogoliubov-Valatin transformation \cite{pitaevskii, peth, book2}, valid for every polarization state $\sigma$, as
\begin{equation}
\hat \alpha_k = u_k \hat b_k + v_k \hat b_{-k}^\dag \, , \quad 
\hat \alpha_{-k}^\dag = u_k \hat b_{-k}^\dag + v_k \hat b_k \, .
\label{6.3b} 
\end{equation}
where the new operators satisfy usual boson commotion relations: $[ \hat \alpha_k , \hat \alpha_{k'}^\dag ] = \delta_{k k'}$, and $[ \hat \alpha_k , \hat \alpha_{k'} ] =[ \hat \alpha_k^\dag , \hat \alpha_{k'}^\dag ] = 0$. The appropriate coefficients $u_k$ and $v_k$, will have to satisfy the hyperbolic relation $u_k^2 - v_k^2 = 1$. In this case, the Hamiltonian $\mathcal{\hat H}_\sigma =\mathcal{\hat H}_{0, \sigma} + \mathcal{\hat H}_{1,\sigma}$ can be written in a diagonalized form, as
\begin{equation}
\mathcal{\hat H}_\sigma = E_{0,\sigma} + \sum'_k \hbar \tilde \omega_k \hat \alpha_k^\dag \hat \alpha_k ,
\label{6.4} 
\end{equation}
where the first term represents the energy of the quasi-particle condensate, as determined by
\begin{equation}
E_{0,\sigma} =  g_\sigma N_{0,\sigma}^2  - \frac{1}{2} \sum_k' \left[ \left( \epsilon_{k,\sigma} + g_{\sigma, k} N_{0,\sigma} \right) - \hbar \tilde \omega_\sigma ({\bf k}) \right] \, , 
\label{6.4b} 
\end{equation}
where $g_\sigma \equiv g_{\sigma, 0}$ and $\tilde \omega ({\bf k})$ is the frequency of the elementary excitations of the condensed gas, satisfying the dispersion relation
\begin{equation}
\hbar \tilde \omega_\sigma ({\bf k}) = \sqrt{\epsilon_{k, \sigma} \left( \epsilon_{k, \sigma} + 2 g_{\sigma, k} N_{0, \sigma} \right)},
\label{6.5} 
\end{equation}
These excitations correspond to sound waves in the quasi-particle gas, not to be confused with the well-known ion-acoustic waves that can be excited in the plasma itself. Noting that for condensed quasi-particles, we should take the limit $k^2 \rightarrow0$, we can use the (polariton) approximation for the plasmon and photon dispersion relations, such that
\begin{equation}
\epsilon_{k, \sigma}=\hbar \sqrt{\omega_p^2+c_\sigma^2k^2}-\mu_\sigma \simeq  \frac{\hbar^2k^2}{2m_\sigma} .
\label{6.5b} 
\end{equation}
Replacing this in Eq. (\ref{6.5}), we obtain
\begin{equation}
\tilde \omega_\sigma^2 = k^2 V_{\sigma,k}^2 + \frac{\hbar^2 k^4}{4 m_\sigma^2} \, .
\label{6.6} 
\end{equation}
Here, we have used the dispersive ($k-$dependent) {\sl quasi-particle sound velocity} $V_{k,\sigma}$, as defined by
\begin{equation}
V_{\sigma, k} = \sqrt{\frac{g_{\sigma, k} N_{0, \sigma}}{m_\sigma}} 
\, .
\label{6.6b} 
\end{equation}
This new class of oscillations is possible due to the weak interaction between the plasma quasi-particles. The latter is mediated by the four-wave mixing process, as described by the coupling parameter $g_\sigma \neq 0$.
Notice that this is different from the second sound velocity, because the second sound involves coherent oscillations of both the background medium density (the electron plasma density, in present case) and the quasi-particle density. Here, in contrast, only the quasi-particle density are supposed to oscillate. 

A possible coupling of these elementary oscillations with other plasma modes should also be investigated. In particular, elementary oscillations of photons could couple with plasmons, and elementary oscillations of plasmons could couple with ion acoustic waves (first sound). For instance, the photon BEC excitations can be coupled to the plasmon modes at the intersection between the two dispersion curves. This occurs when
\begin{equation}
\tilde \omega ({\bf k})^2 = \omega ({\bf k})^2 \equiv \omega_p^2 + k^2 S_e^2,
\label{6.7} 
\end{equation}
or, equivalently, at the wavevector $k_c$ such that
\begin{equation}
k_c^4 = \frac{4 m_{\pm }^2}{\hbar^2} \left[ \omega_p^2 + k_c^2 \left(V_{k,\pm }^2 - S_e ^2 \right) \right]
\label{6.7b} 
\end{equation}
Usually, we have $V_{k,\pm }^2 < S_e ^2$. But, for a cold plasma, we could possibly attain equality $V_{k,\pm }^2 = S_e ^2$, which would lead to
\begin{equation}
k_c^2 = \frac{2 m_{\pm }}{\hbar} \omega_p = 2 \frac{\omega_p^2}{\hbar c^2}
\label{6.7c} 
\end{equation}
We should also notice that, for $k_c^2 \geq \omega_p^2 / S_e^2$, plasma waves are strongly damped by electron Landau damping. Resonant coupling between the BEC excitations and plasmon oscillations could then provide an efficient heating transfer mechanism to cool down and damp the thermal oscillations of the condensed photon gas. 
\begin{figure}[t!]
\includegraphics[scale=0.65]{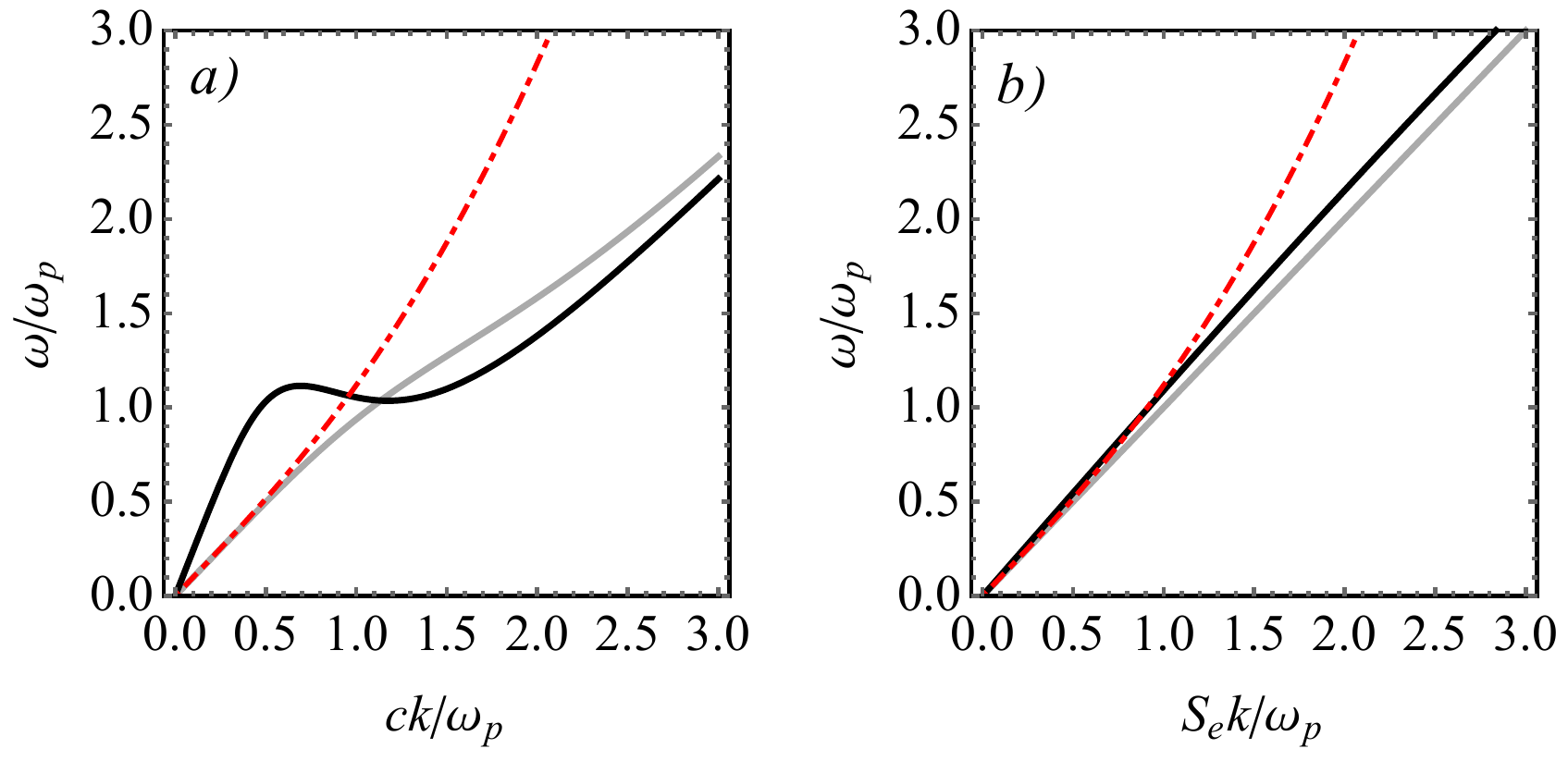}
\caption{(color online) Dispersion relation of the photon (panel a)) and plasmon (photon b)) condensate excitations. In both panels, the lighter solid line corresponds to the weakly relativistic case $\gamma_a \sim 1$ and $m_ec^2\lesssim \hbar \omega_p$, while the darker solid line illustrates the relativistic case $\gamma_a\sim 5$ and $m_ec^2=10 \hbar\omega_p$. A roton minimum appears in the photon dispersion relation as a consequence of the stimulated emission in dense photon gases. The red dashed line is the usual Bogoliubov excitation of dilute atomic gases with contact interaction constante $g=n_0C(0)$.}
\label{fig_Bog}
\end{figure}

\section{Kinetic description of the Compton thermalization}
\label{comptonization}

We now consider the processes leading to thermal equilibrium in a system composed of a fully ionized plasma and a photon gas. As such, we are neglecting eventual absorption and emission of photons from neutral atoms and assume that the photon thermalization is the result of Compton scattering only. The evolution of the photon spectral distribution $N({\bf k})\equiv N_{\sigma}({\bf k})$ is given by the Boltzmann equation \cite{kompaneets_1956, zeldovich_1968}
\begin{equation}
\frac{\partial N({\bf k})}{\partial t}=c\int d \mathbf{p} \int d\Omega \frac{d\sigma}{d\Omega}\left[ f({\bf p}')  N({\bf k}') \left(1+N\left({\bf k}\right)\right)-f({\bf p})N({\bf k})\left(1+N({\bf k}') \right) \right],
\label{eq_BE}
\end{equation}
where $f({\bf p})$ is the electron distribution function. This equation describes energy transfer via Compton scattering events of the form ${\bf p} +\hbar{\bf k}\rightarrow {\bf p}'+ \hbar{\bf k}'$ and vice-versa. The photon scattering rate from momentum $\hbar \mathbf{k}'$ to  $\hbar \mathbf{k}$ is given by the first term of the r.h.s. of Eq. \eqref{eq_BE}. The proportionality to $N({\bf k}')$ and $f({\bf p}')$ is standard in the Boltzmann equation, while the term $\left[ 1+ N({\bf k}')\right]$ accounts for stimulated emission and the Bose-Einstein statistics of the photons.  In the rest frame of the electron, the Compton relation provides
\begin{equation}
\omega_{k'}=\frac{\omega_k}{1+\frac{\hbar \omega_k}{mc^2}\left(1-\cos\theta\right)},
\end{equation}
where $\theta$ is the scattering angle. The differential cross section is given by the Klein-Nishina formula
\begin{equation}
\frac{d\sigma}{d\Omega}=\frac{3}{16\pi}\left(\frac{\omega_{k'}}{\omega_k}\right)^2\left[\frac{\omega_{k'}}{\omega_{k}}+\frac{\omega_k}{\omega_{k'}}-\sin^2\theta \right]\sigma_T,
\end{equation}
where $\sigma_T=8\pi r_c^2/3$ is the Thomson cross section and $r_c=e^2/(4\pi\epsilon_0 mc^2)$ denotes the electron classical radius. A detailed analysis of the evolution of the spectrum $N({\bf k})$ in the presence of successive scatterings off relativistic electrons is difficult, as the energy transfer per scattering is large and involves solving a nonlinear integro-differential equation. Fortunately, we may be interested in the region of the spectrum where photon condensation takes place, i. e. close to the plasma frequency $\omega_p$, such that the soft photon condition $\omega_{k'}-\omega_{k}\ll \omega_p$ may be used, yielding
\begin{equation}
\frac{d\sigma}{d\Omega}\simeq \frac{3}{16 \pi}\left(1+\cos^2 \theta\right)\sigma_T.
\end{equation}
Moreover, the energy transfer is small compared to the electron kinetic energy, $\Delta \equiv \hbar (\omega_{k'}-\omega_{k})/T_e\ll 1$, which for non-relativistic electrons distributed as
\begin{equation}
f(\varepsilon)=\frac{n_0}{\left(2\pi m_e T_e\right)^{3/2}}e^{-\varepsilon/T_e}, \quad \varepsilon=\frac{p^2}{2m_e},
\end{equation}
and energy shift $\varepsilon'\simeq \varepsilon-\Delta T_e$, yields
\begin{equation}
\begin{array}{c}

f(\varepsilon')\simeq \left(1+\Delta+\frac{\Delta^2}{2}\right)f(\varepsilon)\\\\
N(\omega_{k'})\simeq N(\omega_k)+\Delta T_e\partial_\omega N(\omega_k)+\frac{1}{2}\Delta^2 T_e^2\partial^2_\omega N(\omega_k),
\end{array}
\end{equation}
where a change of variables has been used for convenience,  With this approximation, Eq. (\ref{eq_BE}) becomes local in $\mathbf{k}$ (similarly, in $\omega_k$) and reads
\begin{equation}
\frac{\partial N}{\partial t}=\left[N'+N\left(N+1\right)\right]\mathcal{I}_1+\left[\frac{1}{2}N''+\left(N'+\frac{N}{2}\right)\left(1+N\right)\right]\mathcal{I}_2,
\label{eq_BE2}
\end{equation}
where $N\equiv N(x)$ and $N'=\partial_x N(x)$, with $x=\hbar \omega_k/T_e$ and the integrals defined as
\begin{equation}
\mathcal{I}_1=c\int d \mathbf{p}\int d\Omega\frac{d\sigma}{d\Omega}~ \Delta f(\mathbf{p}), \quad \mathcal{I}_2=c\int d \mathbf{p}\int d\Omega\frac{d\sigma}{d\Omega}~\Delta^2 f(\mathbf{p}).
\end{equation}
In order to compute the integrals in Eq. (\ref{eq_BE2}), we must determine $\Delta$ for each individual Compton (or, strictly speaking, Thomson) scattering event. As such, we make use of the energy and conservation laws in the process 
\begin{equation}
\begin{array}{c}
\hbar \omega_k+\varepsilon= \hbar \omega_{k'}+\varepsilon' \\
\hbar {\bf k}+\mathbf{p}=\hbar {\bf k'}+\mathbf{p}',
\end{array}
\end{equation}
Solving for the second expression for ${\bf p}'$ and replacing in the first, we obtain, to first order in $\Delta$ (and neglecting the terms proportional to $\Delta/\omega_k$), the following estimate
\begin{equation}
\hbar \Delta \simeq \frac{\hbar^2}{2m_e}{\bf p}\cdot ({\bf k}-{\bf k}').
\end{equation}
By using the dispersion relation $\omega_k^2=\omega_p^2+c^2k^2$, we then compute the phase-space integrals to order $\mathcal{O}(T_e^2/m_e^2c^4)$ to explicitly obtain
\begin{equation}
\begin{array}{c}
\mathcal{I}_1=\frac{\sigma_Tn_0}{T_e}\frac{\hbar \sqrt{\omega_k^2-\omega_p^2}}{m_e c}\left(4T_e-\hbar \sqrt{\omega_k^2-\omega_p^2}\right)\\\\
\mathcal{I}_2=2\frac{T_e\sigma_Tn_0}{m_ec}\left(\omega_k^2-\omega_p^2\right).
\end{array}
\end{equation}
Finally, by defining the dimensionless variable $\kappa=\hbar \sqrt{\omega_k^2-\omega_p^2}/T_e$ and the so-called Compton time $\tau=(n_0\sigma_T T_e/m_ec^2) t$, we are able to arrive at the  Kompaneets equation \citep{kompaneets_1956}, which is now modified to take into account the fact that photon modes with $\omega_k<\omega_p$ can not propagate in the plasma
\begin{equation}
\frac{\partial N}{\partial\tau}=\frac{f(\kappa)}{\kappa^2}\frac{\partial}{\partial \kappa}\left[\kappa^4\left(\frac{\partial N}{\partial \kappa}f(\kappa)+N+N^2\right)\right],
\label{eq_Kompaneets}
\end{equation}
where $f(\kappa)=\kappa \left(\kappa^2+\hbar^2\omega_p^2/T_e^2\right)^{-1/2}$ is the transformation Jacobian. 
\begin{figure}[t!]
\includegraphics[scale=0.65]{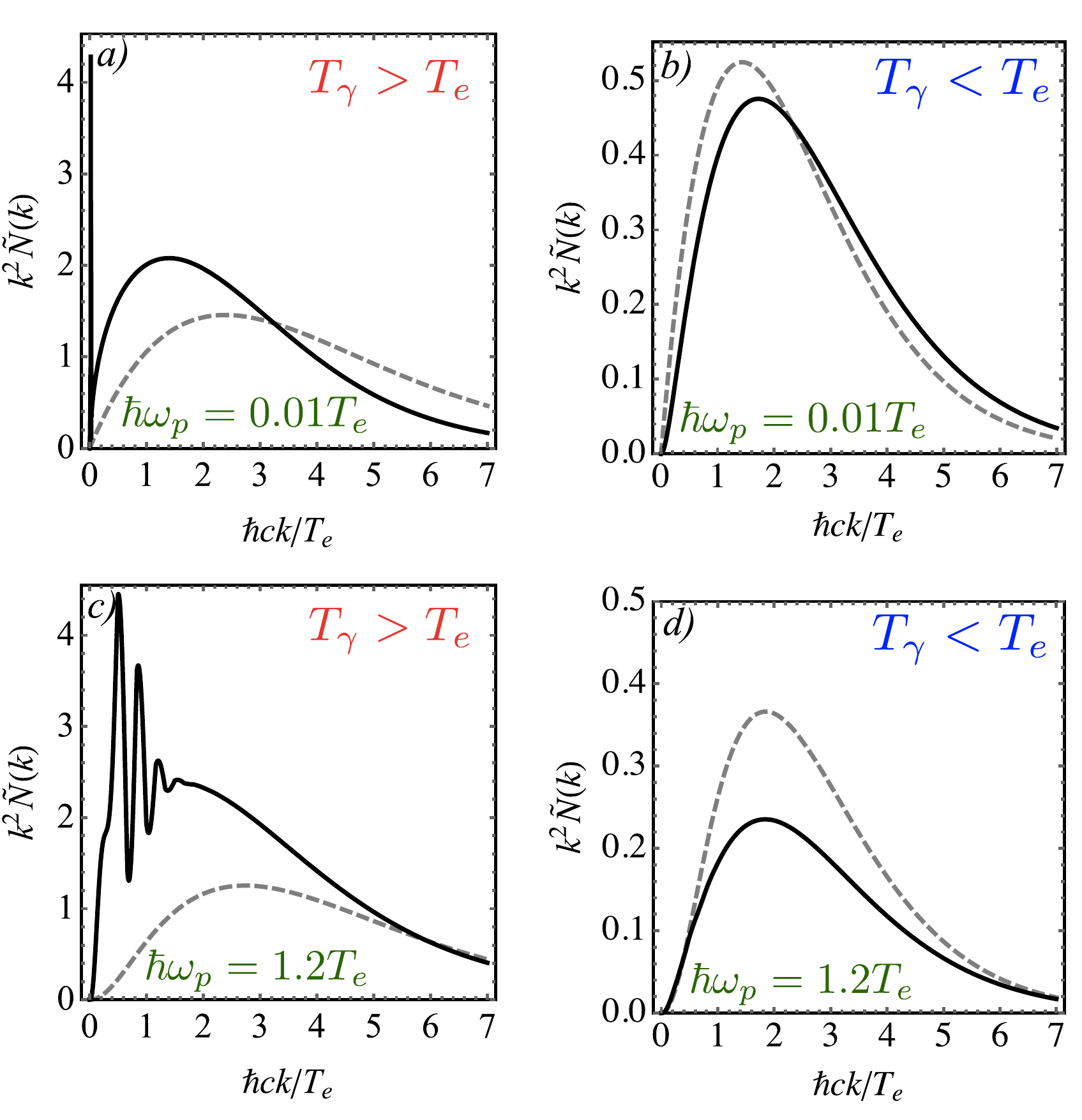}
\caption{(color online) Photon thermalization due to Compton scattering, as resulting from the numerical simulation of the Kompaneets equation for a massive photon of dispersion $\omega^2=\omega_p^2+c^2k^2$. Panels a) and c) (resp. d) and d)) depict the normalized distribution function $k^2\tilde N(k)= k^2 T_e^2N(k)/(\hbar ^2 c^2)$ for the situation where the photon temperature is initially larger (smaller) than the electron temperature in the plasma. We have considered the dilute $\hbar\omega_p\ll T_e$ and the dense $\hbar\omega_p\sim T_e$ for illustration, observing that in the latter situation the inverse Compton scattering leads to the formation of harmonics before the photon condensation.}
\label{fig_thermalization}
\end{figure}
We notice that the steady-state solution to Eq. \eqref{eq_Kompaneets} may not be a Planck distribution, as a consequence of the particle-number conservation in the ``comptonization" process. This can be readily seen by taking the stationary solution $\partial N/\partial \tau=0$, which yields $\partial N/\partial \kappa=-\frac{1}{f(\kappa)}N(N+1)$. For the particular case $f(\kappa)=1$, i.e. in the absence of the plasma, an analytical solution provides the Bose-Einstein distribution $N(\kappa)=1/\left(z e ^\kappa-1\right)$, with $z=e^{-\kappa_0}$ the fugacity. To illustrate the condensation dynamics in the case of a finite value of the plasma frequency, we solve the Eq. \eqref{eq_Kompaneets} numerically. We employ a split-step spectral method and assume the photon distribution to be initially planckian, $N_{\rm initial}=1/(\exp^{(\hbar\omega_k/T_\gamma)}-1)$ \cite{comment}, where $T_\gamma$ is the initial temperature of the photon gas. The results are illustrated in Fig. \ref{fig_thermalization}. As we can observe, for $T_\gamma<T_e$, the distribution with reaches a steady-state corresponding to a Planck distribution at the temperature $T_e$. In contrast, for $T_\gamma >T_e$, inverse Compton scattering provides the accumulation of photons near the mode $k=0$ (i.e. at the bottom of the dispersion $\omega_p$). This effect is accompanied by the formation of a thermal cloud of photons at the temperature $T_e$, thus suggesting that Bose-Einstein condensation of photons, in the thermodynamical sense, may indeed take place. According to our simulations, this process takes place after $\tau \sim 1$, which is of the order of the Compton time. We moreover observe that, when the plasma is sufficiently dense ($\hbar \omega_p\sim T_e$), the photon thermalization is accompanied by the formation of peaks near $\omega_p$, being the most prominent that located at the bottom of the dispersion $k=0$. 

\section{Conclusions}
\label{conclusion}

As we have discussed, transverse and longitudinal photons (in short, photons and plasmons) in a plasma acquire a mass due to their interaction with the electrons. Such an effective mass then results on a finite chemical potential of the photon gas which, given its bosonic statistics, may undergo Bose-Einstein condensation. By establishing a statistical analysis, we have been able to compute this temperature explicitly, observing that it follows the same density dependence $\sim n^{2/3}$ as in more conventional (e.g. alkali atoms) systems. By performing the canonical quantization of the electromagnetic energy of a photon gas in the plasma, we derived an Hamiltonian for the free quasi-particles. Moreover, by considering the four-wave mixing processes in the plasma, and proceeding to the quantization, we introduced an interaction Hamiltonian for the photons in the condensate. This allowed the discussion of the elementary excitations on top of the condensate, which we compared with the Bogoliubov spectrum. Finally, we have derived a kinetic equation that describes the Compton and inverse Compton scattering in the plasma that leads to the thermalization of the photon gas and, eventually, to its condensation. \par
Although our approach involved an unified description of both photons and plasmons, there are substancial aspects that distinguishes them. For example, under certain density and temperature conditions, plasmon (Langmuir) waves may interact with the ion-acoustic waves significantly, what implies taking into account the three-wave mixing processes neglected here. As discussed in the literature, this would result on the collapse of Langmuir waves and, therefore, compromise some of our conclusions. Careful consideration of such effects will most likely deserve our attention in the future. Moreover, the kinematic analysis performed in Sec. VII for photons may be not directly applicable to the case of plasmons. The first modification would be the time scale at which termal equilibrium may be achieved, which is expected to be increased by a factor of $c/S_e$, the ration between the speed of the light and the electron thermal velocity. The second, and more importantly, modification relates the effect of Brillouin scattering, which may impact both the thermalization dynamics and the condensate depletion. Finally, because of the interactions near the bottom of the dispersion are dominated by four-wave mixing, a self-consistent dynamics of the condensate must involve the description of both the thermal cloud, in the spirit of the Zaremba-Nikuni-Griffin model \cite{ZNG, proukakis_2008}, and the coupling to a Boltzmann equation for the plasma reservoir, similar to that of exciton-polariton condensates pumped out-of-resonance \cite{hybrid}. It is our judgment that those important questions should be addressed in future works.

\begin{acknowledgements}

J. T. M. would like to thank the financial support of CNPq Brazil. H. T. acknowledges the Security of Quantum Information Group for the hospitality and for partially providing the working conditions. H. T. also thanks the support from Funda\c{c}\~{a}o para a Ci\^{e}ncia e a Tecnologia (Portugal) through the grant number IF/00433/2015. Stimulating discussions with J. D. Rodrigues are warmly acknowledged. 

\end{acknowledgements}

\bigskip


\begin{thebibliography}{99}
\bibitem{comment} Because the plasma frequency $\omega_p$ is finite, the initial distribution $N_{\rm initial}$ implicitly contains a chemical potential $\mu=\hbar\omega_p$ (neglecting the relativistic $\gamma$ factor, for simplicity). Strictly speaking, we are thus initializing the photons with a distribution of the Bose-Einstein type.
\end{thebibliography}
\end{document}